\documentclass[sigconf,nonacm]{acmart}

\AtBeginDocument{%
  }

\setcopyright{none}
\usepackage{bm}
\usepackage{enumitem}

\usepackage{xspace}
\usepackage{calligra}
\usepackage{amsmath}
\newcommand{\change}{$\mathbb{CHANGE}$\xspace}
\newcommand{\contextualize}{\protect{$\mathbb{C}$}ontextualize\xspace}
\newcommand{\harmonize}{\protect{$\mathbb{H}$}armonize\xspace}
\newcommand{\anticipate}{\protect{$\mathbb{A}$}nticipate\xspace}
\newcommand{\negotiate}{\protect{$\mathbb{N}$}egotiate\xspace}
\newcommand{\generate}{\protect{$\mathbb{G}$}enerate\xspace}
\newcommand{\evolve}{\protect{$\mathbb{E}$}volve\xspace}

\begin{document}

\title{Architecting AgentOps Needs {$\varmathbb{CHANGE}$}}

\author{Shaunak Biswas}
\email{shaunak.biswas@research.iiit.ac.in}
\affiliation{%
  \institution{\textit{SERC, IIIT Hyderabad}}
  \city{Hyderabad}
  \country{India}
}

\author{Hiya Bhatt}
\email{hiya.bhatt@research.iiit.ac.in}
\affiliation{%
  \institution{\textit{SERC, IIIT Hyderabad}}
  \city{Hyderabad}
  \country{India}
}

\author{Karthik Vaidhyanathan}
\email{karthik.vaidhyanathan@iiit.ac.in}
\affiliation{%
  \institution{\textit{SERC, IIIT Hyderabad}}
  \city{Hyderabad}
  \country{India}
}

\begin{abstract}

The emergence of Agentic AI systems has outpaced the architectural thinking required to operate them effectively. These agents differ fundamentally from traditional software: their behavior is not fixed at deployment but continuously shaped by experience, feedback, and context. Applying operational principles inherited from DevOps or MLOps, built for deterministic software and traditional ML systems, assumes that system behavior can be managed through versioning, monitoring, and rollback. This assumption breaks down for Agentic AI systems whose learning trajectories diverge over time. This introduces non-determinism making system reliability a challenge at runtime. We argue that architecting such systems requires a shift from managing control loops to enabling dynamic co-evolution among agents, infrastructure, and human oversight. To guide this shift, we introduce \change, a conceptual framework comprising six capabilities for operationalizing Agentic AI systems: {\contextualize, \harmonize, \anticipate, \negotiate, \generate, and \evolve}. \change provides a foundation for architecting an AgentOps platform to manage the lifecycle of evolving Agentic AI systems, illustrated through a customer-support system scenario. In doing so, \change redefines software architecture for an era where adaptation to uncertainty and continuous evolution are inherent properties of the system.

\end{abstract}




\maketitle

\section{Introduction}
Engineering AI systems has progressively evolved over the past few years. 
The software industry is experiencing an unprecedented surge in the deployment of Agentic AI systems composed of autonomous agents. Large Language Models (LLMs), which are foundational models trained on vast corpora \cite{llm_def, ipek_llm}, now serve as the reasoning core of AI agents (hereafter referred to as agents) that interpret complex situations and act through external tools and services \cite{liu2025agent,agents_cain_sreemaee}. These agents are increasingly used in domains such as customer support, software development, and decision support \cite{maad, agents_cain_sreemaee}. Their behavior evolves during operation as they respond to feedback and changing conditions. Yet one foundational problem remains unresolved: \textit{we do not know how to architect operations for systems that think, learn, and evolve.}

Operational thinking inspired from MLOps for Machine Learning Enabled Systems (MLS) assumes behavior can be governed through versioning, monitoring, and rollback. That assumption fails once agents continue to learn after release. Versioning an initial prompt or configuration captures a starting point, not the evolving internal state that drives current behavior. Conventional metrics such as latency, cost, or accuracy provide little traction on how an agent’s reasoning changes through experience. Guardrails that predefine allowable actions help with obvious errors, yet they often fail to guide behavior in novel situations that were never specified at design time leading to an {\em unknown unknown} scenario~\cite{calinescu2020understanding,hezavehi2021uncertainty}.

When multiple agents interact within an Agentic AI system, these weaknesses compound. Reported improvements over single-agent baselines are frequently small, and failures recur in recognizable forms such as ambiguous specifications, misaligned objectives across agents, and insufficient verification of plans and tool use. The consistent pattern is clear: many breakdowns arise from architectural choices about representation, coordination, and oversight rather than from raw model capacity \cite{multi_agent_sys_fail}. Incremental fixes rarely change this outcome. These issues are not tooling gaps. Continuous learning introduces non-determinism at runtime, which means reliability cannot be certified once at deployment and forgotten. It must be managed continuously through representations and mechanisms that track how agents acquire experience, how they align with one another and with human intent, and how they are guided when behavior begins to drift. Addressing this challenge demands architectures that treat Agentic AI systems, their supporting infrastructure, and human operators as co-evolving components of a single adaptive system.

\noindent\textbf{Our Position}: To make this shift, we introduce \change, a conceptual framework for designing and governing AgentOps across the full lifecycle. \change defines six essential capabilities that an AgentOps platform must realize: \contextualize, for representing an agent’s experiential state; \harmonize, for maintaining alignment in multi-agent settings; \anticipate, for predicting and preparing for unanticipated situations; \negotiate, for dynamically managing autonomy and oversight; \generate, for enabling agents to propose and validate new tools or capabilities; and \evolve, for assessing and governing long-term adaptation across the system.

We illustrate these capabilities through the running example of an Agentic AI system with two customer-support agents, \textit{Alice} and \textit{Bob}. Their independent learning and ocassional conflicts show how \change addresses such challenges. The rest of the paper explains the framework and shows how it helps Agentic AI systems stay reliable and aligned as they evolve.
\section{Related Works}
Operationalizing intelligent systems such as traditional ML models, LLMs, and AI agents presents a growing set of architectural challenges \cite{genai_taibi, mlops_challenges,llmops_challenges,he2025llm,liu2025agent}. The operationalization of ML-enabled systems has been addressed through \textit{MLOps} \cite{amershi2019software}, which establishes standardized processes for developing, deploying, and maintaining ML models. MLOps frameworks emphasize automation, version control, continuous integration, and lifecycle management, making them effective for task-specific models whose objectives and data distributions remain relatively stable \cite{zarour2025mlops}. However, the emergence of foundation models, like LLMs, has complicated this picture. These models are not trained for a single downstream task but instead serve as general-purpose reasoning engines that can be adapted through prompting, fine-tuning, or in-context learning \cite{xi2025rise}. Traditional MLOps pipelines, designed for static model-dataset pairs, are ill-equipped to manage such flexibility. This gap has led to the rise of \textit{LLMOps} \cite{llmops_challenges_check}, which extends MLOps principles to handle foundation models by managing prompt templates, fine-tuning strategies, deployment configurations, and performance monitoring.
% \footnote{LLMOps has been described in several ways; here we adopt the definition of John et al.~\cite{} as the systematic process of managing, evaluating, and deploying foundation models across diverse applications.}
LLMOps thus focuses on ensuring the reliable operation of powerful but general-purpose models that can be adapted to a wide range of downstream tasks. While LLMOps represents a step forward in handling foundation models, it primarily treats the model as a component within a managed pipeline. It does not account for systems where the model itself acts autonomously. In practice, many modern AI systems now embed LLMs within agents that perceive their environment, plan actions, and execute tasks through external tools or APIs \cite{xi2025rise,wang2025agents, react}. These agents use LLMs to reason, and make decisions by combining model outputs with memory, feedback, and goal-driven behavior \cite{he2025llm,liu2025agent}. Once deployed, such agents operate continuously, learning from interaction rather than retraining, and adapting their strategies based on accumulated experience \cite{sapkota2025ai}. This shift from managing static or prompt-controlled models to operating continuously learning, autonomous entities introduces new operational demands. Pipelines built for MLOps and LLMOps assume bounded behavior, static evaluation criteria, and clear deployment checkpoints. In contrast, Agentic AI systems evolve dynamically, modifying their internal state, generating new context, and sometimes diverging from their initial specifications \cite{vaidhyanathan2025software,liu2025agent, multi_agent_sys_fail}. Existing operational paradigms provide little guidance for governing this evolution, aligning multiple interacting agents, or ensuring traceability across behavioral change. To address this emerging gap, we propose the \change framework which introduces six capabilities: \contextualize, \harmonize, \anticipate, \negotiate, \generate, and \evolve. These enable AgentOps platforms \cite{dong2024agentops} to manage an Agentic AI system where agents reason, adapt, and learn continuously. By grounding operational practices in these capabilities, the \change framework moves beyond static model management toward the sustained, interpretable, and coordinated operation of autonomous AI agents.

\section{Running Example}
\label{Running_example}
Consider an Agentic AI system deployed by a global e-commerce platform. This system includes two AI agents, \textit{Alice} and \textit{Bob}. \textit{Alice}, a customer support agent, manages refund requests and delivery complaints. Alice begins with a well-defined system prompt, company policy documentation, and a set of tools for order lookup and refund processing. Over time, she handles thousands of customer interactions, learning patterns in how users phrase problems and when exceptions are likely to occur. Occasionally, she infers solutions not explicitly covered by policy, offering partial refunds or creative goodwill gestures that improve satisfaction but deviate from approved guidelines. \textit{Bob}, a logistics support agent, manages logistics coordination and occasionally collaborates with Alice when return shipments are delayed. Unlike a traditional ML system where a model produces predictions without changing after deployment, an Agentic AI system involves agents that interact with their environment and adapt their behaviour over time. As both agents adapt independently, subtle behavioral differences emerge: Bob prioritizes efficiency, while Alice emphasizes empathy. Their recommendations begin to conflict in overlapping cases leading to increased response latency, inconsistent user experience and reduced reliability of the overall service workflow. Meanwhile, customer expectations and company rules evolve, forcing both agents to operate outside their original competence. This affects the system's ability to maintain consistent behavior over time, raising concerns about long-term maintainability and operational stability.

These interactions expose several fundamental challenges in architecting operations for AgentOps platform in managing the Agentic AI system:

\begin{table}[h!]
\centering
\scriptsize
\caption{Relationship between Identified Challenges and Corresponding \change Capabilities}
\label{tab:challenges_mapping}
\renewcommand{\arraystretch}{1.5}
\setlength{\tabcolsep}{3pt}
\begin{tabular}{p{0.4cm}p{5.27cm}p{2.2cm}}
\toprule
\textbf{ID} & \textbf{Challenge Description} & \textbf{\change Capability} \\
\midrule
\textbf{CH1} & Each agent keeps learning from its environment, which changes how it understands and acts over time. The challenge is to understand its current reasoning and anticipate its future actions. & \textit{\contextualize}, \textit{\anticipate} \\

\textbf{CH2} & When several agents work together, each evolves differently. The challenge is to keep them coordinated while allowing them to adjust their adaptation boundaries through negotiation with other agents and human supervisors. & \textit{\harmonize}, \textit{\negotiate} \\

\textbf{CH3} & As the Agentic AI system evolves, agents may acquire new abilities or retire old ones. The challenge is to manage these changes so that the system continues to improve without losing stability. & \textit{\generate}, \textit{\evolve} \\
\bottomrule
\end{tabular}
\label{tab:change_challenges}
\end{table}

In the remainder of this paper, we revisit the example of Alice and Bob and use the \change framework to address these challenges.

\begin{figure*}[t]

  \centering
  % If the PDF has multiple pages, pick one with page=<n>
  \includegraphics[width=\textwidth]{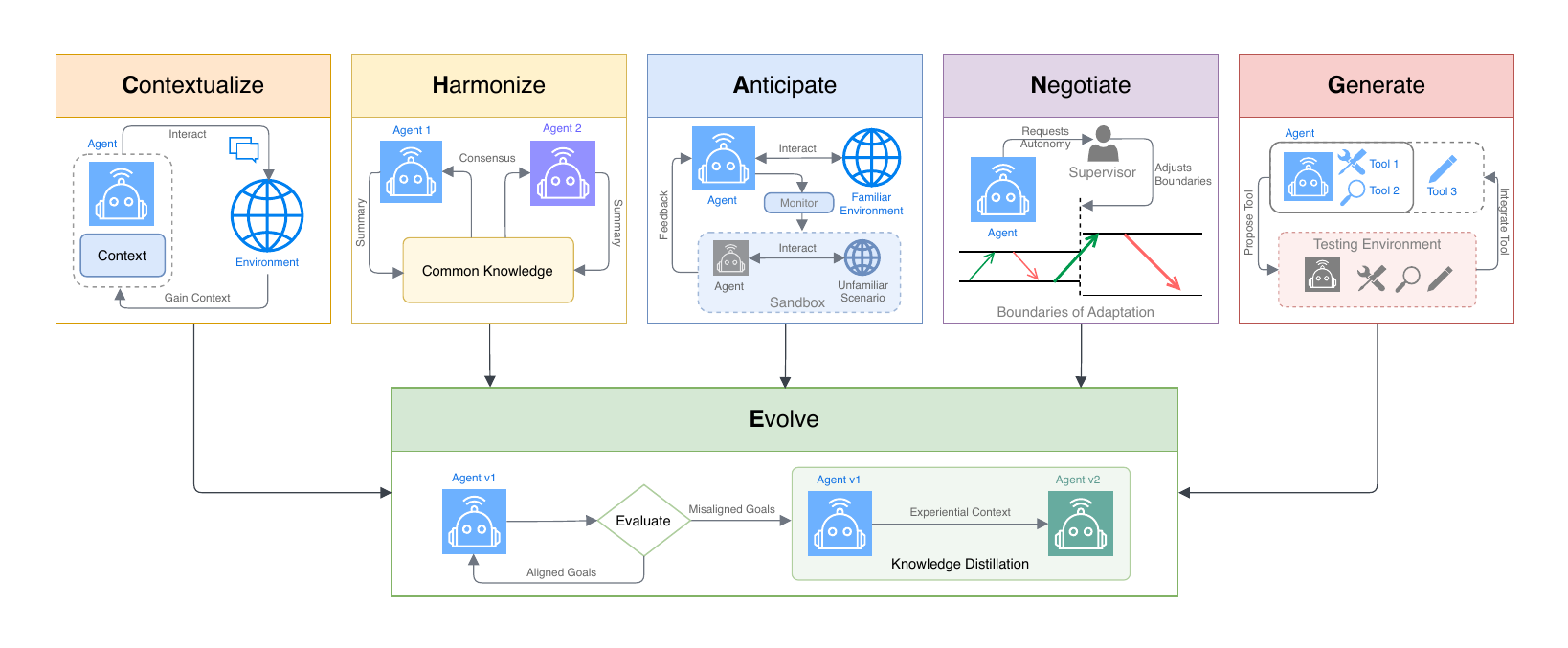}
  \caption{The \change framework illustrated through agent interaction scenarios}
  \Description{Block diagram of the \change framework showing the main components and data flow.}
  \label{change_figure}
\end{figure*}
\section{The {$\varmathbb{CHANGE}$} Framework}
The \change framework defines six capabilities that an AgentOps platform requires to operationalize an Agentic AI system effectively: \contextualize, \harmonize, \anticipate, \negotiate, and \generate, all contributing to \evolve. Each capability addresses a distinct limitation in current operational practices, and together they enable the AgentOps platform to collaborate with agent autonomy rather than constrain it.

\subsection{\contextualize}
\noindent \textbf{CH1} in Table \ref{tab:change_challenges} highlights that an agent’s behavior evolves as it gains experience. This makes its state dependent on accumulated interactions rather than static configuration files or checkpoints. \contextualize addresses this by requiring the AgentOps platform to represent the agent’s experiential history: what it has learned, when it learned it, and how this knowledge influences later decisions. Work in continual and lifelong learning indicates that retaining experience can help maintain a balance between \textit{plasticity} (learning new information) and \textit{stability} (preserving prior knowledge) \cite{plasticity_neurips}. By encoding an agent’s learning trajectory, \contextualize enables stable, interpretable, and long-term adaptation.

This capability operates through a continuous feedback loop, as shown in Figure~\ref{change_figure}. The agent interacts with its environment, generating new context that is captured and integrated into its internal state. This can be realized through \textit{context-aware versioning}, where the agent’s cognitive state is periodically stored as a versioned snapshot, and through structured state representations that record learned associations and reasoning traces. Recent work on rule-based explainability \cite{neurips_contextualize} demonstrates how decision traces can be organized into structured, human-interpretable rules; similar techniques could inform how contextual information is encoded and queried at runtime.

Going back to our running example (Section \ref{Running_example}), Alice is no longer the agent defined by her initial prompt; her thousands of customer interactions have created a unique experiential context. An architecture guided by the \contextualize would not treat her policy deviations as anomalies but as data points in an evolving history. It would manage her entire experiential context as a versioned artifact, allowing operators to understand precisely how and why her empathetic strategies emerged by replaying key segments of her operational history.
\subsection{\harmonize}
\textbf{CH2} in Table \ref{tab:challenges_mapping} notes that when several agents work together, each evolves differently, making it difficult to maintain alignment within the Agentic AI system. \harmonize partly addresses this challenge by ensuring that the interactions among multiple agents remain coordinated as they learn and adapt. Its purpose is not to enforce uniform behavior, but to maintain shared understanding so that the system behaves coherently despite internal diversity.

This capability builds on research in distributed consensus and cooperative multi-agent systems. Studies on goal alignment in reinforcement learning \cite{aaai_goal_alignment} and shared-memory coordination \cite{neurips_shared_memory} show that agents can maintain collective coherence by periodically exchanging compact representations of their internal state. Extending this to LLM-based systems, recent work \cite{ijcai_llm_influence} demonstrates that architectural design itself can shape how agents influence each other’s reasoning. Building on these insights, \harmonize can be implemented as a \textit{behavioral consensus protocol}, where agents share summarized reasoning traces in a shared \textit{common knowledge} space, as shown in Figure~\ref{change_figure}. This shared context allows divergences in goals or strategies to be detected early and alignment to emerge through continual interaction rather than centralized control.

In the running example (Section~\ref{Running_example}), \harmonize addresses the growing divergence between Alice and Bob. As Alice optimizes for empathy and Bob for efficiency, their independent learning leads to conflicting recommendations in shared logistics and support cases. Through the behavioral consensus protocol, the system detects these inconsistencies by comparing their summarized reasoning stored in the shared \textit{common knowledge} space. It then triggers a consensus step where Alice and Bob align on a common policy for handling delayed returns. This coordination preserves their individual strengths while ensuring that their combined behavior remains consistent with system-level objectives.
\subsection{\anticipate}
\textbf{CH1} in Table \ref{tab:challenges_mapping} highlights that as agents continuously learn from their environment, their reasoning and actions evolve in ways that are difficult to predict. \anticipate addresses this challenge by enabling Agentic AI systems to forecast how an agent’s behavior is likely to change under new or shifting conditions (behavioral drift). This form of drift is different from model or data drift in traditional ML systems as it emerges from an agent's ongoing interaction rather than from shifts in training data. Instead of reacting after problems occur, it prepares the system to act in advance by identifying early signs of behavioral drift and developing suitable responses before undesired outcomes arise.

As shown in Figure~\ref{change_figure}, this capability operates through a \textit{feedback} process linking real-world monitoring and simulated experimentation. The agent interacts with the familiar environment while a monitoring layer tracks its decisions and learning patterns. When potential drift is detected, a mirrored version of the agent is instantiated in a controlled \textit{sandbox} that reproduces operational conditions. This sandbox may be implemented using a digital-twin-like environment where the agent can be exposed to unfamiliar or extreme scenarios. Since agents exhibit inherently non-deterministic behavior, relying on deterministic testing is insufficient. Scenario-based testing in the sandbox enables systematic exploration of multiple possible outcomes. Insights from this simulated environment are then fed back to the live system to refine policies or adaptation strategies without interrupting ongoing operations. Prior work on digital twin-driven predictive maintenance \cite{digit_hiya} shows that integrating real-time monitoring with virtual experimentation enables early fault detection and proactive adaptation. This improves system reliability and operational stability.

In the running example (Section~\ref{Running_example}), \anticipate helps detect early signs of Alice’s behavioral drift. As she learns to handle customer complaints more empathetically, her behavior gradually diverges from company policy. The system maintains a predictive model of her decisions and deploys her simulated counterpart in a sandbox environment that replicates key properties of the real system. This setup allows the system to observe her reaction to challenging or unfamiliar refund cases and intervene early through policy reminders, targeted retraining, or human feedback. This ensures that her learning continues to improve customer satisfaction while remaining within operational guidelines.

\subsection{\negotiate}
\textbf{CH2} in Table \ref{tab:challenges_mapping} highlights that as agents evolve within an Agentic AI system, coordination requires not only shared understanding, but 
also controlled flexibility in how their adaptation boundaries are managed. These boundaries represent the acceptable limits within which a system can operate safely \cite{ilias_adaptation_intent}. \negotiate addresses this challenge by providing a structured mechanism through which agents can request changes to their autonomy, while supervisors decide whether these adjustments are justified based on trust, performance, and risk.

As shown in Figure~\ref{change_figure}, agents operate within defined adaptation boundaries but can request changes when new situations arise. These requests are routed to a supervisor, who reviews the agent’s reasoning and decides whether to approve, reject, or modify the proposed change. This supervised pathway ensures that autonomy expands only when the agent demonstrates reliability in specific contexts. Alternatively, the system can autonomously adjust boundaries through mechanisms such as control-theoretic approaches \cite{harmone_hiya} or learning based methods \cite{competence}, allowing continuous adaptation without direct human intervention. Both approaches enable flexible management of autonomy: one through negotiated oversight, the other through self-adaptation \cite{kephart2003vision}. Together, they ensure that adaptation remains safe, traceable, and aligned with system-level objectives.

In the running example (Section~\ref{Running_example}), \negotiate governs how Alice manages autonomy when faced with decisions outside of the standard policy. Let us suppose she encounters a complex refund case involving a long-term customer that does not fit any predefined rule. Instead of immediately escalating the issue, the architecture allows her to request temporary permission to handle the case directly, providing an explanation for her reasoning. The supervisor reviews this request and decides whether to approve or modify it. If similar cases are consistently resolved well, Alice’s adaptation boundary can expand to include such scenarios in the future. Through repeated successful cases, the system learns when it can safely extend an agent’s autonomy.

\subsection{\generate}

\noindent \textbf{CH3} in Table \ref{tab:challenges_mapping}  highlights that as an Agentic AI system evolves, its constituent agents may encounter situations that existing tools or methods cannot handle. The challenge is to enable them to extend system capabilities safely without disrupting overall stability. \generate addresses this by allowing agents to identify gaps in their abilities, propose new tools or approaches, and validate them before integration. Rather than depending entirely on human designers, this capability enables systems to grow by incorporating agent-driven innovations that emerge through real operational experience.

As shown in Figure~\ref{change_figure}, \generate introduces a structured pathway for agent-driven innovation. When an agent detects a missing capability or identifies a more efficient approach, it can \textit{propose} a new tool or process. This proposal is then evaluated in a controlled \textit{testing environment}, where the agent can safely experiment without affecting live operations. If the tool performs reliably and meets defined safety and performance criteria, it is \textit{integrated} into the operational system and becomes part of the agent’s available toolkit. This pipeline ensures that new ideas are tested, verified, and gradually absorbed into the system, allowing it to evolve without compromising stability.

In the running example (Section~\ref{Running_example}), \generate is relevant when Alice frequently depends on Bob from logistics to retrieve detailed tracking information for international orders. Over time, she notices that this repeated exchange increases latency and introduces unnecessary dependencies. Drawing on her operational experience, Alice proposes a new capability: an automated query tool that consolidates shipment data from multiple carriers. As shown in Figure~\ref{change_figure}, this proposal is first tested in the testing environment to verify accuracy and reliability before being integrated into the live system. Once validated, the new tool becomes available to all agents, removing the performance bottleneck arising from the coordination between Alice and Bob. 

\subsection{\evolve}

\textbf{CH3} in Table \ref{tab:challenges_mapping} highlights that as systems grow, agents may create new abilities or phase out old ones. The challenge lies in managing these changes safely so that the system continues to improve. For an Agentic AI system managed by an AgentOps platform with the preceding \change capabilities, evolution becomes a natural consequence. \contextualize captures an agent’s experiential history, \harmonize maintains coordination among agents, \anticipate forecasts behavioral drift, \negotiate regulates adaptation boundaries, and \generate introduces new capabilities. Together, these mechanisms create a continuously learning ecosystem where agents adapt and improve through interaction. The role of \evolve is to govern this process, ensuring that adaptation remains aligned with the system’s long-term objectives.

As shown in Figure~\ref{change_figure}, \evolve introduces a structured process for evaluating and managing an agent’s lifecycle. Each evolved version of an agent is periodically \textit{evaluated} to check whether its goals remain aligned with the system’s objectives. When alignment is maintained, the agent continues to operate and refine its behavior through feedback. When goals become misaligned, \evolve enables the Agentic AI system to initiate a transition phase. In this phase, the current agent’s experiential context is extracted and transferred through a process of \textit{knowledge distillation} \cite{distill}, where the useful representations, reasoning traces, and learned associations from the older agent (Agent v1) are selectively distilled into a new successor (Agent v2). This ensures that the system retains valuable operational experience while eliminating inefficiencies or outdated behavioral patterns. Through this process of evaluation and distillation, \evolve ensures that adaptation becomes an inherent capability of the Agentic AI system.

This process also extends to the running example in Section~\ref{Running_example}. Over time, Alice’s empathetic strategies, once beneficial, begin to conflict with a new organizational policy emphasizing consistency over personalization. The system detects that her evolved reasoning no longer aligns with these updated goals. Rather than discarding her progress, a transfer process is initiated: a new agent, “Alex,” is instantiated with updated base policies, while selected elements of Alice’s experiential context, such as her ability to interpret nuanced language and resolve ambiguous cases, are distilled into Alex’s training corpus. Alice is then formally phased out, with her knowledge preserved and repurposed. In this way, \evolve ensures that the system continues to improve through successive generations of agents.
\newline

\noindent Architecting AgentOps platforms guided by the \change framework requires accepting that evolution is inevitable. The goal for such systems is not to prevent change, but to manage it. As Agentic AI systems learn and adapt via experience it becomes essential to address whether the evolution is desirable or not, and when to constrain it. Therefore architectures must include mechanisms to evaluate alignment between an agent’s evolving behavior and system objectives. \textbf{Empirical validation} of the \change framework could proceed through controlled simulations using established multi-agent platforms \cite{wu2024autogen}. Such simulations would enable comparative analysis between Agentic AI systems operationalized through  existing AgentOps platforms versus those incorporating \change capabilities. Key metrics could include behavioral alignment divergence over time (testing \harmonize), the accuracy of behavioral drift prediction (testing \anticipate), the rate of successful boundary negotiations (testing \negotiate), and the overall system resilience to environmental shifts, ultimately assessing the effectiveness of co-evolutionary design. While concrete metrics for evaluating the full lifecycle of evolving agents are still emerging, we view \change as a foundation for architecting AgentOps platforms, enabling self-adaptation in Agentic AI systems, and guiding how such systems should be observed and improved in practice.

\section{Conclusion}

Agentic AI systems introduce operational demands that current software architectures are not designed to accommodate. Traditional operational practices, such as MLOps, assume that system behaviour can be predicted at design time. This breaks down in Agentic AI systems, where learning continues after deployment and adaptation becomes an inherent property of operation. The \change framework responds to this shift by defining six capabilities that an AgentOps platform must realise to operate Agentic AI systems. We showed how \contextualize, \harmonize, \anticipate, \negotiate, \generate, and \evolve work together to support AgentOps. Using the example of \textit{Alice} and \textit{Bob}, we illustrated how these capabilities help represent experiential state, maintain alignment across agents, prepare for behavioural drift, manage autonomy with appropriate oversight, support agent-driven improvements, and guide the long-term evolution of the Agentic AI system. The goal is not to stop agents from changing, but to make change observable and well governed. The \change framework does not remove uncertainty since it is inherent in AI enabled systems and even more pronounced in Agentic AI systems.  Instead it provides structure for working with it. Agents will still encounter unfamiliar situations, and not every proposed capability will be suitable for deployment. What \change offers is a way to detect such situations early and respond in a measured and informed way.

\textbf{{A Call to AI Engineering Community: }}The rise of Agentic AI systems demand a shift in how we think about building and operating software. These systems learn from experience, adapt their behaviour over time, and display traits that resemble human decision making. If we expect such systems to operate responsibly, we must confront fundamental questions:
\begin{enumerate}[label=\roman*.]
    \item \textit{What skills and tools do software architects require to design and monitor Agentic AI systems responsibly?}
    \item \textit{How do we make architectural design decisions when system behaviour cannot be fully specified at design time and continues to change during operation?}
    \item \textit{How can we develop evaluation methods to assess such non-deterministic systems?}
\end{enumerate}

Moving AgentOps forward will require shared practices, clear evaluation criteria, and evidence from real deployments.

% Embracing this challenge requires a clear position on the nature of control. This is not a call for reckless deployment of uncontrolled agents. Rather, it is recognition that effective control comes from understanding and shaping behavioral evolution, not from imposing rigid constraints that agents will inevitably circumvent or that will prevent agents from realizing their potential. The CHANGE framework presented in this paper offers the conceptual foundations for this shift. Each of its pillars addresses fundamental limitations in current AgentOps and offers principles for building systems that work with agent autonomy rather than against it.

The \change framework is a step toward operating Agentic AI systems in a way that recognises adaptation as a natural part of their lifecycle. Control comes from understanding how behaviour evolves and guiding it with the right checks, rather than restricting it through rigid rules. The aim is not to stop agents from changing, but to make that change visible and well governed. The \change framework provides an initial foundation for this shift. We encourage the community to refine and extend these ideas so that AgentOps can mature into a disciplined and evidence based practice grounded in shared methods and real deployments. The agents are already evolving. The question is whether our architectures will evolve with them.

\bibliographystyle{ACM-Reference-Format}
\bibliography{references}
\end{document}